\documentclass{aa}  

\usepackage{graphicx}
\usepackage{txfonts}
\usepackage[english]{babel}
\usepackage{amssymb}
\usepackage{amsmath}
\usepackage{mathrsfs}
\usepackage{latexsym}
\usepackage{longtable}
\usepackage{multirow}
\usepackage{epsf}
\usepackage{color}
\usepackage{float}
\usepackage{enumerate}
\usepackage{txfonts}
\usepackage{natbib}
\usepackage{comment}
\usepackage{capt-of}
\usepackage{rotating}
\usepackage{ulem}

\newcommand{\hd}{HD\,139614 }

\begin{document} 

   \title{Shadowing and multiple rings\\ in the protoplanetary disk of HD\,139614
\thanks{Based on observations performed with SPHERE/VLT under program ID 096.C-0248(B) and  099.C-0147(B).}}

\author{G.\,A.\,Muro-Arena\inst{1}
\and M.~Benisty\inst{2,3}
\and C.~Ginski\inst{1}
\and C.~Dominik\inst{1}
\and S.~Facchini\inst{4}
\and M.~Villenave\inst{2,5}
\and R.~van Boekel\inst{6}
\and G.~Chauvin\inst{2,3}
\and A.~Garufi\inst{7}
\and T.~Henning\inst{6}
\and M.~Janson\inst{8}
\and M.~Keppler\inst{6}
\and A.~Matter\inst{9}
\and F.~M\'enard\inst{2}
\and T.~Stolker\inst{10}
\and A.~Zurlo\inst{11,12,13}
\and P.~Blanchard\inst{13}
\and D.~Maurel\inst{2}
\and O.~Moeller-Nilsson\inst{6}
\and C.~Petit\inst{14}
\and A.~Roux\inst{2}
\and A.~Sevin\inst{15}
\and F.~Wildi\inst{16}
}

\institute{Anton Pannekoek Institute for Astronomy, University of Amsterdam, Science Park 904,1098XH Amsterdam, The Netherlands \email{g.a.muroarena@uva.nl}
\and Univ. Grenoble Alpes, CNRS, IPAG, 38000 Grenoble, France
\and Unidad Mixta Internacional Franco-Chilena de Astronom\'{i}a (CNRS, UMI 3386), Departamento de Astronom\'{i}a, Universidad de
Chile, Camino El Observatorio 1515, Las Condes, Santiago, Chile
\and European Southern Observatory, Karl-Schwarzschild-Str. 2, 85748 Garching, Germany
\and European Southern Observatory, Alonso de C\'ordova 3107, Vitacura, Casilla 19001, Santiago, Chile
\and Max Planck Institute for Astronomy, K\"{o}nigstuhl 17, 69117 Heidelberg, Germany
\and INAF, Osservatorio Astrofisico di Arcetri, Largo Enrico Fermi 5, I-50125 Firenze, Italy
\and Department of Astronomy, Stockholm University, Stockholm, Sweden
\and Laboratoire Lagrange, Universit\'e  Côte d’Azur, Observatoire de la  Côte d’Azur, CNRS, Boulevard de l’Observatoire, CS 34229, 06304 Nice Cedex 4, France
\and Institute for Particle Physics and Astrophysics, ETH Zurich, Wolfgang-Pauli-Strasse 27, 8093 Zurich, Switzerland
\and N\'ucleo de Astronom\'ia, Facultad de Ingenier\'ia y Ciencias, Universidad Diego Portales, Av. Ejercito 441, Santiago, Chile
\and Escuela de Ingenier\'ia Industrial, Facultad de Ingenier\'ia y Ciencias, Universidad Diego Portales, Av. Ejercito 441, Santiago, Chile 
\and Aix Marseille Universit\'e, CNRS, LAM - Laboratoire d'Astrophysique de Marseille, UMR 7326, 13388, Marseille, France 
\and DOTA, ONERA, Université Paris Saclay, F-91123, Palaiseau France 
\and LESIA, Observatoire de Paris, Universit\'e PSL, CNRS, Sorbonne Universit\'e, Univ. Paris Diderot, Sorbonne Paris Cit\'e, 5 place Jules Janssen, 92195 Meudon, France
\and Geneva Observatory, University of Geneva, Chemin des Mailettes 51, 1290 Versoix, Switzerland
}

\date{}

  \abstract
    {Shadows in scattered light images of protoplanetary disks are a common feature and support the presence of warps or misalignments between disk regions. These warps are possibly due to an inclined (sub-)stellar companion embedded in the disk.}
    {We study the morphology of the protoplanetary disk around the Herbig\,Ae star HD\,139614 based on the first scattered light observations of this disk, which we model with the radiative transfer code \texttt{MCMax3D}.}
    {We obtained $J$- and $H$-band observations in polarized scattered light with VLT/SPHERE that show strong azimuthal asymmetries. In the outer disk,  beyond $\sim$30\,au, a broad shadow spans a range of $\sim$240 deg in position angle, in the East. A bright ring at $\sim$16\,au also shows an azimuthally asymmetric brightness, with the faintest side roughly coincidental with the brightest region of the outer disk. Additionally, two arcs are detected at $\sim$34\,au and  $\sim$50\,au. We created a simple 4-zone approximation to a warped disk model of HD\,139614 in order to qualitatively reproduce these features.  The location and misalignment of the disk components were constrained from the shape and location of the shadows they cast.}
    {We find that the shadow on the outer disk covers a range of position angle too wide to be explained by a single inner misaligned component. Our model requires a minimum of two separate misaligned zones –  or a continuously warped region – to cast this broad shadow on the outer disk. A small misalignment of $\sim$4$^{\circ}$ between adjacent components can reproduce most of the observed shadow features.}
    {Multiple misaligned disk zones, potentially mimicing a warp, can explain the observed broad shadows in the HD\,139614 disk. A planetary mass companion in the disk, located on an inclined orbit, could be responsible for such a feature and for the dust depleted gap responsible for a dip in the SED.}

\titlerunning{Shadowing and multiple surface rings\\ in the protoplanetary disk HD\,139614}
\authorrunning{Muro-Arena et al.}
\maketitle

\section{Introduction}
\label{sec:introduction}
One of the first steps towards planet formation is the coagulation of small dust grains to form planetesimals. 
However, a known barrier in planetesimal formation is the fast radial drift of dust pebbles driven by the gas sub-Keplerian rotation, on timescales shorter than the ones required for significant grain growth to occur \citep{weidenschilling1977}.
A way to overcome that issue is to trap dust grains in local pressure maxima \citep{klahr1997}, that can be generated in various ways such as planet-disk interactions \citep{rice2006,pinilla2012}, a change in material properties \citep[ice line,][]{stammler2017}, a dead zone \citep{dzyurkevich2010, pinilla2016}, allowing them to grow efficiently. Dust traps can be located at any radius, leading to localized over-densities of large (mm-sized) grains, which would appear as concentric rings in images of the thermal emission at mm wavelengths and might be necessary to explain the global spectral indices measured in various star forming regions \citep{ricci2010}. Azimuthally asymmetric dust traps, caused by vortices and possibly at the edge of the depleted inner cavity, can also produce significant growth of dust material \citep[e.g.,][]{birnstiel2013,casassus2015}. 

As dust trapping appears to be a key process to form planets and leads to clear observational signatures, the quest for substructures in protoplanetary disks was strongly motivated. In the past few years, two observing techniques reaching high angular resolution have led to stunning disk images showing a wide diversity of structures. In the thermal dust continuum emission, as probed by the Atacama Large Millimeter Array (ALMA), many disks show annular substructures (rings and gaps) at very different radii, with a wide range of ring/gap intensity contrast \citep[e.g.,][]{jane2018a, long2018}. In particular, the rings detected by the DSHARP campaign, on 20 bright disks, \citep{andrews2018} are narrower than the pressure scale height, supporting that they trace
dust trapping \citep{dullemond2018}. So far, only three disks around single young stars show spiral arms in the thermal continuum \citep{jane2018b}, which can result from gravitational instability \citep{kratter2016}, or gravitational interaction with a companion \citep{bae2018}. Other disks show in addition to rings a localized asymmetric feature \citep{cazzoletti2018,perez2018}, often referred to as dust crescent, spread over a small range of position angles, whose origin can also triggered by planet-disk interactions \citep{zhang2018}. Transition disks (disks with a dust depleted inner cavity, or/and a clear dip in the spectral energy distribution at mid-IR wavelengths) also show evidence for small substructures, such as multiple rings, beyond the cavity \citep[e.g.,][]{dong2018a, perezS2019, keppler2019}.

Another, complementary, way to search for and characterize substructures in disks is through scattered light high contrast imaging \citep[e.g.,][]{garufi2018}. If the disk is significantly resolved, and high contrast is achieved by the observations, fine structures are revealed as bright rings, spiral arms, broad cavities and dark localized regions. In contrast to thermal emission that traces the bulk dust material in the midplane, scattered light imaging traces the tenuous dusty surface layers of the disk, where the stellar light is reflected and polarized by small (sub-micron or micron-sized) dust grains. It therefore directly traces the stellar irradiation pattern onto the disk and is very sensitive to any departure from a smooth morphology of the disk \citep{facchini2018, nealon2019}. Interestingly, the features detected in scattered light and in thermal emission often do not have a direct correspondence; for example, some disks show a clear $m=2$ spiral arm pattern in scattered light, while they appear as rings with lopsided asymmetries in the ALMA images \citep{cazzoletti2018, uyama2018, kraus2017}. 

As scattered light imaging is sensitive to the stellar irradiation, it allows one to search for misalignments between various disk regions. While studying the morphology of the innermost disk region is challenging due to its very small radial extent, often marginally resolvable by optical interferometry \citep{lazareff2017}, scattered light imaging of the outer disk can indirectly reveal the presence of a misaligned inner disk. In this scenario, depending on the misalignment angle, the outer disk image will show narrow shadow lanes \citep[e.g.,][]{pinilla2015, stolker2016, benisty2017, casassus2018}, broad extended shadows  \citep{benisty2018} or low amplitude azimuthal variations \citep{debes2017,Poteet2018}. In some cases, studies of the CO line kinematics support a misalignment between inner and outer disk regions \citep{loomis2017,perez2018}. The exact origin of such a misalignment is still unclear. In the case of T Tauri stars, if the stellar magnetic field is inclined, it can warp the innermost edge of the disk, which would then rotate at the stellar period \citep[AA\,Tau;][]{bouvier2007}. Alternatively, inner and outer disk regions can have different orientations if the primordial envelope had a different angular momentum vector orientation at the time of the inner/outer disk formation \citep{Bate2018}. Other scenarios involve the presence of a massive companion/planet that is inclined with respect to the disk. If the companion is massive enough, the disk can break into two separate inner and outer disk regions, that can then precess differently and result in an important misalignment between each other  \citep[e.g.,][]{nixon2012b,facchini2013,nealon2018,zhu2019}. A clear example of such a scenario is the disk around HD\,142527, in which an M-star companion was detected \citep{biller2012}, likely on an inclined and eccentric orbit \citep{lacour2016, claudi2019}. Dedicated hydrodynamical simulations successfully reproduce most of the observed features in this disk \citep[eccentric cavity, spiral arms, misaligned inner disk and shadows;][]{price2018}.

In this paper we focus on the protoplanetary disk around the young intermediate mass star HD\,139614, located at a distance of $\sim$135\,pc \citep{gaia2018}. This A7V spectral type star has an estimated mass of $1.5\pm0.1$ M$_{\odot}$ and an age of $15.6\pm4$ Myr \citep{fairlamb2015}, and it is not known to have any binary companions. The spectral energy distribution (SED) shows a dip at mid infrared (IR) wavelengths, evidence for a dust depleted inner region. The inner disk was resolved by the Very Large Telescope Interferometer (VLTI) at near and mid-IR wavelengths \citep{matter2014}, indicating the presence of dust in the innermost region. These observations, supported by radiative transfer models of the SED and visibilities, allowed to infer the presence of a gap between 2.5 and 6\,astronomical units (au) \citep{matter-2016}. This result was confirmed by an IR spectroscopic study of the inner disk by VLT/CRIRES \citep{carmona2017}, that showed that the gas is depleted at radii inside $\sim$5\,au although the system still shows on-going accretion  \citep[$\sim2\times10^{-8}$\,M$_{\odot}$/yr;][]{fairlamb2015}. 
These results support the presence of a companion inside the innermost few au of the disk. In the following, we present the first scattered light observations of \hd obtained with the Very Large Telescope (VLT) Spectro-Polarimetric High-contrast Exoplanet REsearch (SPHERE) instrument \citep{beuzit2019sphere}. Our observations trace the small (sub- and micron-sized) dust grains  at the surface layers of the disk and show a number of features (rings and shadows) that can be qualitatively well-reproduced by a radiative transfer model of a warped disk.  Our paper is organized as follows: in Sect.\,\ref{sec:datared} we present our observations and the data reduction; in Sect.\,\ref{sec:images} we describe the scattered light images, in Sect.\,\ref{sec:modeling}, the radiative transfer model, and in Sect.\,\ref{sec:discussion} we discuss our findings.

\begin{figure*}[!h]
\center
\includegraphics[width=0.9\textwidth]{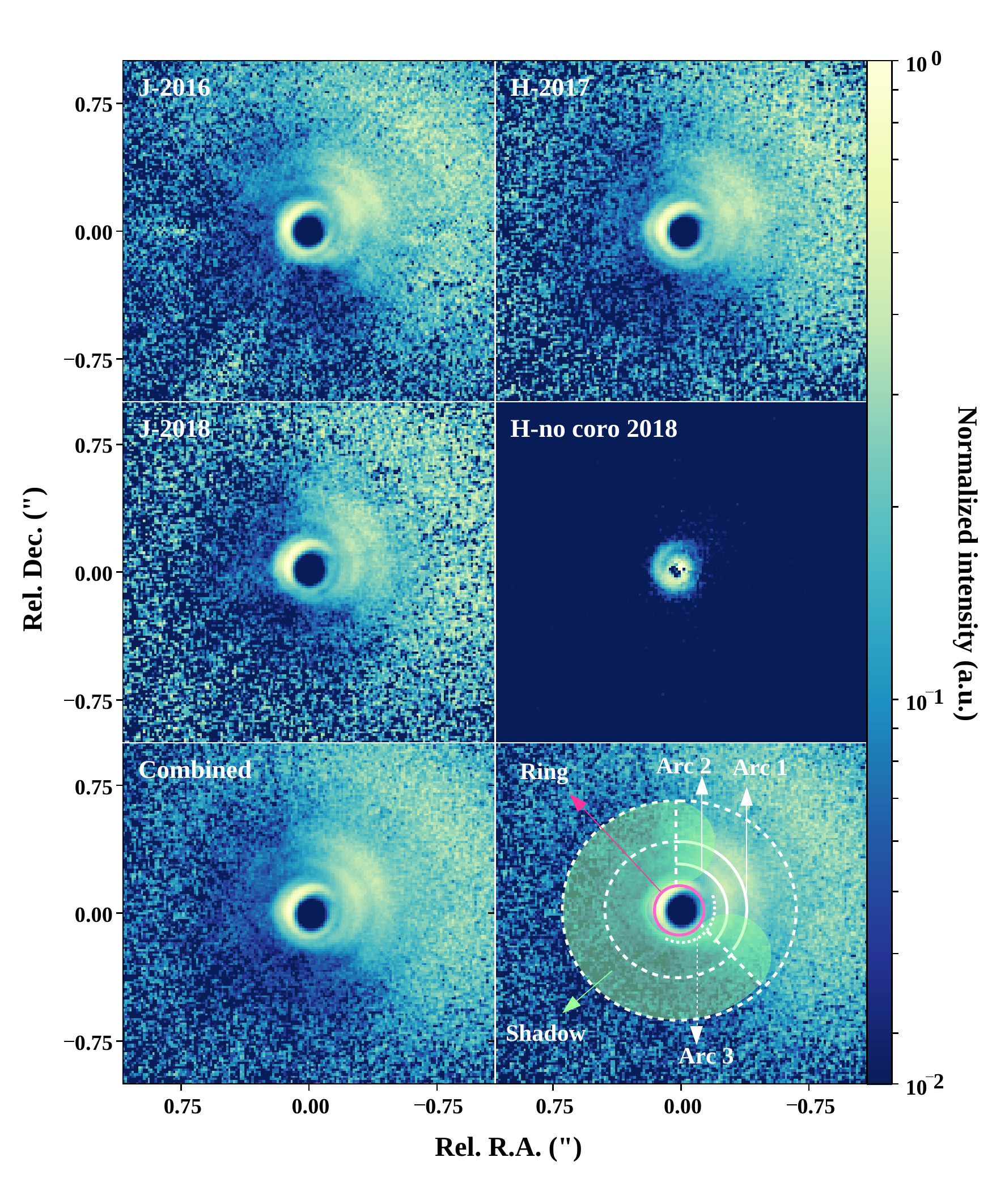} 
\caption{SPHERE/VLT scattered light observations of HD\,139614, obtained at four different epochs (top four panels). The top two panels and middle left panel images were obtained with a coronagraphic mask. The middle right panel shows the non-coronagraphic image. The labels indicate the wavelength range and year of the observations. The bottom row shows the combined image of all coronagraphic images, with an annotated version pointing out the main features in the right column. All images are normalized to their peak intensity after r$^{2}$-scaling, with the exception of the non-coronographic epoch, which is not r$^{2}$-scaled since this only accentuates the noise at large radii.} 
\label{fig:all_epochs}
\end{figure*}

\section{Observations and data reduction}
\label{sec:datared}

The HD\,139614 system was observed on four different occasions between March 2016 and June 2018. 
We give a detailed summary of the observation dates and modes in Table~\ref{tab: obs_summary}.
All observations were obtained using the differential polarimetric imaging mode (DPI, \citealt{langlois2014}) of the InfraRed Dual-band Imager and Spectrograph (IRDIS, \citealt{dohlen2008}).
With polarimetric differential imaging \citep[PDI; e.g.,][]{kuhn2001,apai2004} one measures the linear polarization of the light scattered by dust grains in the disk. This technique enables us to efficiently remove the unpolarized stellar contribution and to image with high contrast the outer disk from which we detect polarized scattered light.
In three of the epochs we used an apodized Lyot coronagraph with a diameter of 185\,mas (\citealt{martinez2009,carbillet2011}), while in one epoch we specifically targeted the smallest resolvable separations without coronagraph.
A half-wave plate was used to modulate the light and measure different Stokes parameters. 
All observations were executed such that one full polarimetric cycle took approximately 4\,min to limit changes in instrumental polarization during the cycle. 
Total integration times vary by epoch between 16\,min (for the non-coronagraphic observation) and 59.7\,min. 
Observations were executed in J and H-band using the IRDIS broad band filters ($\lambda_0$=1.258, $\Delta\lambda$=0.197\,$\mu$m; $\lambda_0$=1.625, $\Delta\lambda$=0.290\,$\mu$m). For the non-coronagraphic observations in June 2018 we used in addition a neutral density filter that reduced the incoming flux by a factor of 10. This was done to prevent saturation and to ensure the smallest possible inner working angle. 

The last observation epochs, performed in 2018, include detector dithering to reduce the effect of bad pixel and flat fielding errors (this mode was only offered recently for DPI observations).
Observing conditions were fair during the coronagraphic observations with average seeing values in the visible between 0.8\,arcsec and 0.9\,arcsec and reasonable atmosphere coherence time above 2.4\,ms.
The conditions were worse for the non-coronagraphic observations with a seeing of 1.1\,arcsec and a short coherence time of 1.5\,ms.
However, even in these conditions a reasonable adaptive optics performance was achieved with several diffraction rings of the stellar point spread function (PSF) visible.

\begin{table*}
 \centering
 \caption{Observation setup and observing conditions.}
  \begin{tabular}{@{}lcccccc@{}}
  \hline 
 Date  		& Filter	& Coronagraph	& DIT [s] & \# Frames & Seeing [arcsec] & $\tau_0$ [ms]\\
 \hline 
 31-03-2016	& BB\_J		& YJH\_ALC	& 64      & 56		&	0.8	&	2.4	\\
 16-05-2017	& BB\_H		& YJH\_ALC	& 64	  & 32		&	0.9	&	-	\\
 06-06-2018	& BB\_H + ND1	& none		& 2	  & 480		&	1.1	&	1.5	\\
 21-06-2018	& BB\_J		& YJH\_ALC	& 64	  & 32		&	0.9	&	4.5	\\
 
 \hline

\hline\end{tabular}
\label{tab: obs_summary}
\end{table*}

The data reduction of the coronagraphic data followed the double difference approach \citep{kuhn2001} and is described in detail in \cite{deboer2016}, thus we only give a short summary.
After initial reduction, including flat-field, sky subtraction and bad pixel masking, the star was centered in all frames using dedicated center calibration frames with a symmetrical waffle pattern inserted by the AO system as reference.
In a next step we performed the polarimetric double difference by subtracting for each frame the simultaneously taken orthogonal polarized images and then for each Stokes parameter the two images taken with 45\,deg half-wave plate rotation.
These steps were performed for each polarimetric cycle independently, then the final Stokes Q and U images of all cycles were collapsed.
In these images we then performed an instrumental polarization correction as outlined by \cite{canovas2011}. We used the bright AO correction radius as reference region to determine the correction factor.

For the non-coronagraphic observation we implemented an extra step to center the images accurately. We fitted a Moffat function to the stellar PSF in each individual frame and re-centered them based on the fit.
In these images, we also used a different region for the instrumental polarization correction since the AO correction radius has very low flux due to the short exposure times. 
In this case we used the inner 5 pixels of the stellar PSF as a reference. 
Since this is within one resolution element, the azimuthal polarization signal will not be resolved anymore and thus we expect close to zero linear polarization given the low inclination of the detected disk.

After the final instrumental polarization corrected Stokes Q and U images were produced for the coronagraphic and the non-coronagraphic data sets, we calculated the radial Stokes parameters Q$\phi$ and U$\phi$ by the formulas given in \cite{schmid2006} and \cite{avenhaus2014}.
Q$\phi$ contains all the azimuthally polarized signal as positive values and potentially radially polarized signal as negative values. U$\phi$ contains all signal that is polarized with an angle of 45\,deg relative to azimuthal or radial direction.
For a near face-on disk, as is the case for the HD\,139614 system, U$\phi$ is not expected to contain any astrophysical signal and can thus be used as convenient noise estimator.  

Finally, in order to create the highest signal-to-noise image with the available data, we also produced an image in which we combined the three coronagraphic observation epochs.
For this purpose we used the final Q$\phi$ images for each epoch. We then scaled the pixel scale of the H-band epoch with a factor of 0.99934 to account for the small difference in pixel scales between the two sets of broad band filters that we used \citep{maire2016}.
Since all images had the same individual frame exposure time, we did not apply a normalization to the individual images before we finally median combined them.

\section{Scattered light images} 
\label{sec:images}
Figure\,\ref{fig:all_epochs} presents the scattered light images obtained at the four different epochs, three with a coronagraph and one without. We also show a composite image, obtained after combining all three coronagraphic epochs as described in Sect.\,\ref{sec:datared}. All three coronagraphic images appear quite similar: at radii close to the masked region ($\sim$0.12\arcsec{} or ~16 au), the disk appears as a bright ring with enhanced brightness towards the East/North-East. Beyond a radius of $\sim$0.22\arcsec{} or $\sim$30 au, the outer disk shows a striking azimuthal asymmetry with a brighter region located in the West/North-West, between position angles (PAs) of ~240$^{\circ}$ and 360$^{\circ}$. The outer disk can be seen to extend in scattered light out to a radius of $\sim$1.5\arcsec{} or $\sim$200 au. 

To quantify the asymmetry, we compute radially averaged azimuthal profiles of the disk, using apertures of 1 pixel in radius, after deprojecting the image using a disk inclination of 17.6$^{\circ}\pm 3.1^{\circ}$ and PA of 276.5$^{\circ}\pm 3.1^{\circ}$. These values of inclination and position angles were obtained by fitting a razor thin Keplerian disk model to the archival ALMA $^{13}$CO moment 1 map (program ID 2015.1.01600.S;  beam=0.72\arcsec{}$\times$0.52\arcsec{}). 

\begin{figure}[t]
\includegraphics[width=0.48\textwidth]{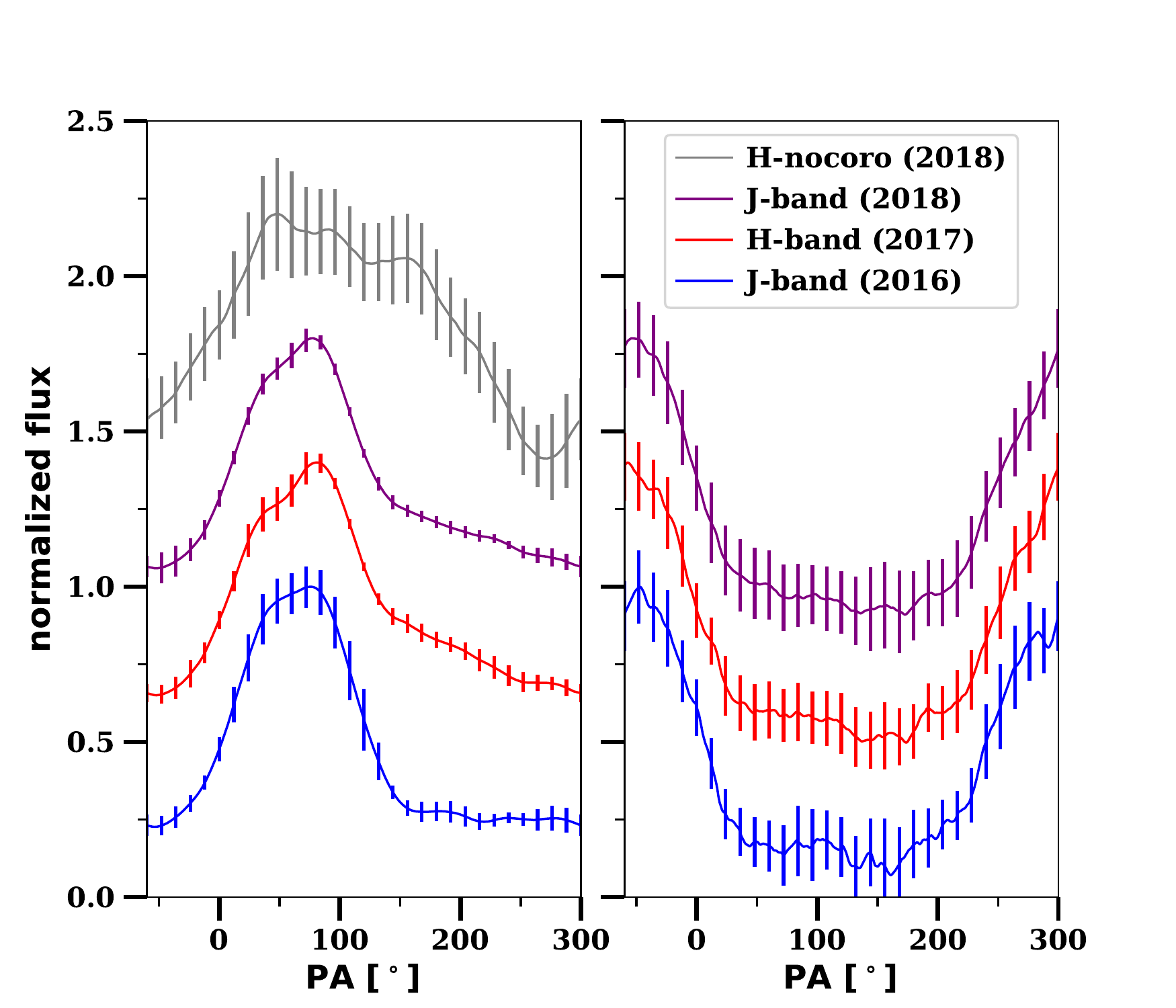} 
\caption{\textit{Left:} Normalized azimuthal profiles of the ring from all three coronographic and one non-coronographic epochs, averaged over radii of $\sim$0.1-0.15 arcsec, with a vertical offset of 0.4 between profiles. \textit{Right:} Normalized azimuthal profiles of the outer disk from all three coronographic epochs, averaged over radii of $\sim$0.23-0.8 arcsec, with a vertical offset of 0.4 between profiles. In both panels, the sequence of curves from bottom to top is the same as in the legend.} 
\label{fig:profiles_az_all}
\end{figure}

For the outer disk, we consider a region between radii of 0.25\arcsec{} and 1.0\arcsec{}. As shown in Figure\,\ref{fig:profiles_az_all} (right panel),  between ~0 and 240$^{\circ}$, the disk signal is fainter than the peak by a factor ~4. We found that the shape of the radially-averaged azimuthal profiles for each epoch can be well fitted by a Gaussian function centered at a PA of $\sim$300$^{\circ}$ and with a FWHM of $\sim$100$^{\circ}$. 

The bright ring, located at ~0.12”, right outside of the region covered by the coronagraph, also presents an azimuthal asymmetry, with peak brightness around a PA of ~60-70$^{\circ}$. The azimuthal profile of this ring was measured in the same way as for the outer disk (Fig.\,\ref{fig:profiles_az_all}, left panel). The faintest region of the ring is fainter than the peak brightness by a factor ~4-5. The shape of the profile appears to vary slightly between the different epochs, though this is likely due slight variations in the centering of the images due to the normal stellar jitter ($\sim$1-2 mas or $\sim$0.1 pixels) behind the coronograph. In all coronographic epochs, as well as in the single non-coronographic epoch, the faintest side of the ring is the West side, roughly coincidental with the region where the outer disk (r>30\,au) is brightest.  

\begin{figure}[t]
\includegraphics[width=0.48\textwidth]{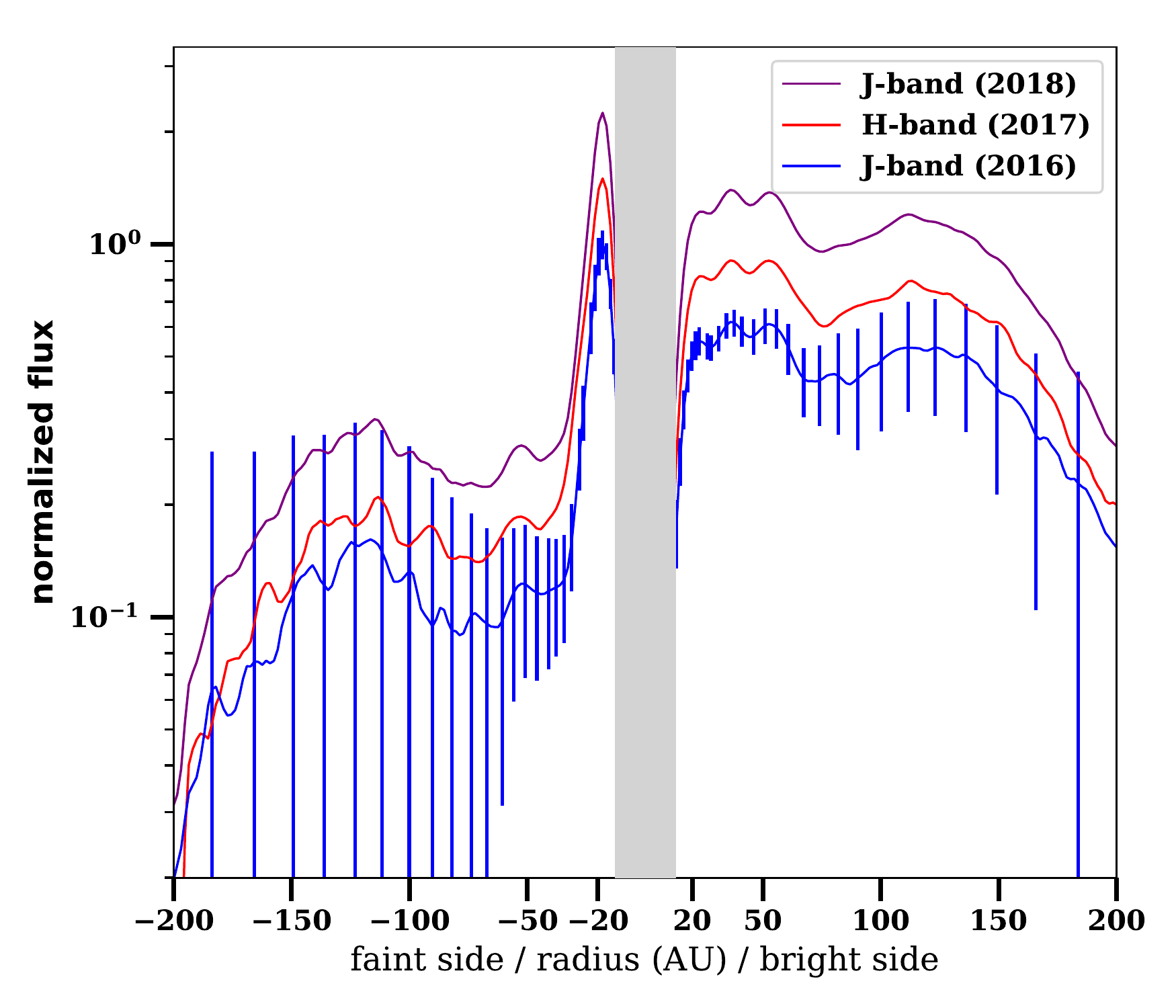} 
\caption{Azimuthally averaged radial profile of the disk, measured after $r^{2}$-scaling of the images: negative radii correspond to the East (faint outer disk) side of the image, with the profile averaged over position angles of 20-220$^{\circ}$; positive radii correspond to the West (bright outer disk) side of the image, with the profile averaged over position angles of 240-360$^{\circ}$. The grey shaded area shows the region covered by the coronograph. Profiles has been offset vertically by a factor 1.5, and the errorbars shown for the J-band 2016 data are representative for all three epochs.}
\label{fig:profiles_rad}
\end{figure}

Three bright arcs are visible in the brighter region of the outer disk, at radii of $\sim$0.37\arcsec (50\,au, Arc\,1), $\sim$0.25\arcsec{} (34\,au, Arc\,2)  and $\sim$0.18\arcsec{} (24\,au, Arc\,3). Arc\,3 is faint and seen most clearly in the combined image. It is, however, also detected in the azimuthally-averaged radial profile of the outer disk (Fig.\,\ref{fig:profiles_rad}), along with Arcs 1 and 2. The figure additionally shows that Arc\,1 is also detected in the radial profiles of all three epochs along the fainter side of the outer disk. At a radius of $\sim$0.73\arcsec{} (or $\sim$100\,au) the brightness of the outer disk appears to increase slightly along the fainter side of the disk (PA $\sim$0-240$^{\circ}$). This is seen more clearly in the composite image. 

The 2018 H-band non-coronagraphic image (Figure \ref{fig:all_epochs}, second row, right) shows what appears to be a crescent-shaped gap or shadow inside the bright ring at 16\,au. Inside this 'gap', we detect scattered light from an inner disk component.

\section{Modeling}
\label{sec:modeling}
\subsection{Motivation}
To understand the features described in Section\,\ref{sec:images} we aimed to construct a three-dimensional model of the disk using Monte Carlo radiative transfer code \texttt{MCMax3D} \citep{min2009}. For this goal we chose to focus on three of the disk’s most prominent features: (1)~the azimuthal brightness asymmetry observed in the outer disk at radii r$\gtrsim$30\,au, (2)~the asymmetry in the bright ring immediately outside the coronagraph (r$\sim$10-20\,au), and (3)~the faint arcs seen in the outer disk at r$\sim$34 and 50\,au. 

{\bf Polarization and scattering effects.} Such features do not naturally arise in a model of a fully coplanar disk with a power-law scale height and surface density profiles. While the angular dependence of polarization efficiency and scattering phase functions can result in asymmetries in disks moderately to highly inclined  with respect to our line of sight, this does not seem to be the main source of brightness asymmetry in our data, considering the low inclination of the disk. The polarized phase function is relatively flat at this inclination, and previous observations of inclined disks show that even for moderate inclinations the asymmetry produced by the phase function is not large enough to completely obscure one side of the disk \citep{stolker2016}. Additionally, the asymmetry arising from the polarized phase function should be symmetrical along the major axis of the disk. At a position angle of $\sim$276.5$^{\circ}$ this translates roughly to an East-West symmetry, in direct contrast to what is seen in our data. The fact that the ring at $\sim$16\,au and the outer disk beyond $\sim$30\,au are asymmetrical in different directions further supports the idea that these asymmetries are not a product of the polarized phase function. The azimuthal asymmetries found in our data therefore require the dust density distribution to be asymmetric (either due to azimuthal asymmetries in the scale height or in the surface density of the disk), or a misalignment between the disk components leading to an asymmetric shadowing of the disk at larger radii.

{\bf A misaligned inner disk.} Shadows caused by misaligned disk components are relatively common and must be long-lived as evidenced by the frequency with which they are observed in scattered light images of protoplanetary disks \citep[e.g.,][]{pinilla2015,stolker2016}. When the misalignment between the inner and outer components is large, twin narrow shadows are cast on the surface of the outer disk \citep{avenhaus2014,marino2015}. For small misalignments, however, when the relative inclination between components is only of a few degrees, a single broad shadow is cast \citep{juhasz2017,debes2017,benisty2018}. The shape of the asymmetry of the outer disk, with the fainter region of the disk appearing roughly constrained to a constant range of PAs between $\sim$0$^{\circ}$ and 240$^{\circ}$ (see Fig.\,\ref{fig:polar_projection}) between radii of $\sim$30 and 100\,au, strongly suggests that the asymmetry is caused by shadowing of the outer disk by an inner disk (r<30\,au) component. This is further supported by the increase in brightness of the outer disk at radii greater than $\sim$100\,au, with the scattering surface of the outer disk pulling up above the shadow at larger radii as the disk aspect ratio increases. A variable scale height of the ring in the azimuthal direction, as opposed to a misalignment of the ring with respect to the outer disk, could also cast a shadow and simultaneously explain both the ring and outer disk asymmetries. This geometry however would be difficult to explain through a physical mechanism, especially over such a broad range of position angles. 

\begin{figure}[t]
\center
\includegraphics[width=0.5\textwidth]{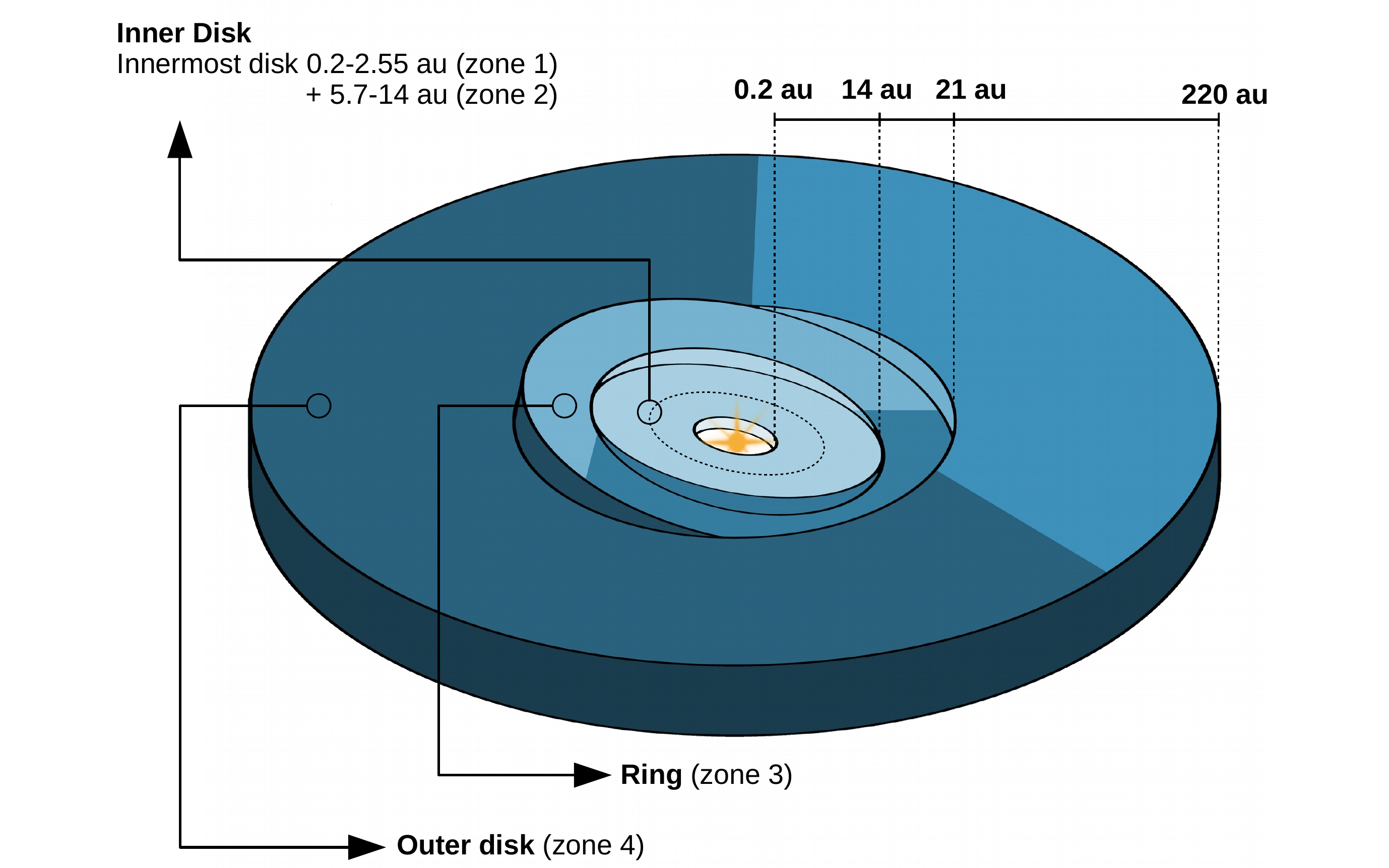} 
\caption{
Sketch of our model, showing the various regions of the disk misaligned with respect to each other. The darker regions on each zone represent the shadowed regions. \textbf{Zones 1 and 2 have the same alignment in our model, and are therefore shown here jointly as the region labeled Inner Disk.}}
\label{fig:geometry}
\end{figure}

{\bf Multiple misaligned components.} The azimuthal asymmetry of the bright ring however suggests that a more complex model might be needed to explain the scattered light observations. If this asymmetry is also caused by shadowing, another misaligned component inside a radius of $\sim$14\,au is required to cast a shadow on the fainter side of the ring. This scenario is summarized in the sketch of Fig.\,\ref{fig:geometry}. This model has the potential of solving one of the issues that show up in models with only two misaligned components, namely, that the shadow cast upon the outer disk  spans about 240 degrees azimuthally, much broader than can be cast by the sole misalignment of the ring. Having this additional inner (r<14\,au) component casting a shadow on both the ring and the outer disk can explain the extent of the shadow of the outer disk as the combination of the shadows cast by both inner components in different directions. 

{\bf Arcs in the outer disk.} The faint arcs seen in the outer disk at $\sim$34 and 50\,au can be caused by either radial variations in the surface density or scale height profiles of the disk. As scattered light images are most sensitive to variations in the slope of the scattering surface \citep{juhasz2015}, with a larger slope locally translating to an increase in brightness, we chose to model the arcs as variations to the radial scale height profile of the disk.


\begin{table}
 \centering
 \caption{Inner and outer radii and disk masses for each of the zones in our model.}
  \begin{tabular}{@{}lccc@{}}
  \hline 
 Zone  		& $R_{in}$	& $R_{out}$	& Mass ($M_{\odot}$)\\
 \hline 
Zone 1 & 0.2 & 2.55 & $8.90\times10^{-11}$ \\
Zone 2 & 5.7 & 14 & $7.65\times10^{-6}$ \\
Zone 3 & 14 & 21 &  $6.45\times10^{-6}$\\
Zone 4 & 21 & 220 &  $1.83\times10^{-4}$\\
 \hline

\hline\end{tabular}
\label{tab:radii}
\end{table}


\subsection{Radiative transfer model}

As a starting point, we consider the radiative transfer model of \citet{matter-2016}, constrained with near- and mid-infrared interferometric observations of HD\,139614 and its SED. This model consists of a dust-depleted inner disk (8.9$\times$10$^{-11}$\,M$_{\odot}$) and a more massive outer disk (1.3$\times$10$^{-4}$\,M$_{\odot}$) separated by a gap between 2.55 and 5.7\,au. The stellar parameters used for our model correspond to those from \citet{matter-2016}, corrected for the Gaia DR2 distance of 134.8\,pc: T$_{\rm{eff}}$= 7850\,K, M= 1.6\,M$_{\odot}$, R= 1.54\,R$_{\odot}$ and the resulting luminosity L= 8\,L$_{\odot}$. 
Our model differs from the one of \citet{matter-2016} in the following: we consider the DIANA standard grain composition \citep{woitke2016} in the entire disk and do not consider a rounded-off rim for the inner edge of the outer disk at 5.7\,au.  We extend the outer disk out to 220\,au to match the extent of the disk as seen in our observations, in particular in the composite H- and J-band combined image. The same surface density profile was maintained for this outer region, and the outer disk mass was increased accordingly to 1.975$\times10^{-4}\,M_{\odot}$.  While it matches the VLTI observations and the SED well, the original  model \citep{matter-2016} is azimuthally symmetric and therefore  fails to reproduce the azimuthal asymmetries of our scattered light images (outer disk, asymmetry of bright ring, and arcs). 
This motivates the changes that we made to this initial model in order to better explain the features seen in the polarized scattered light images in the different epochs.

Our model consists of four zones: an inner disk (zone 1, as in \citet{matter-2016}) and three zones into which the outer disk is divided (zones 2-4) in order to produce the azimuthal asymmetries seen in our SPHERE observations. The radial extent of these zones were constrained from our images: zone 2 corresponds to the component inside $\sim$14\,au  observed in the non-coronagraphic J-band image, zone 3 corresponds to the bright ring at ~16 au, and zone 4 corresponds to the asymmetrical outer disk. The radial extent of each of these zones as well as their disk masses is summarized in Table~\ref{tab:radii}. No gaps were introduced between the three outer disk zones, but the gap between zones 1 and 2 from the \citet{matter-2016} model was kept. 
We note that these radii are only loosely constrained from our data, and are determined by eye from both the coronographic and non-
coronographic data. We ran a grid of models with different inclinations and position angles for zones 1-2 (note that we systematically use the same inclination and position angle for these two zones) and for zone 3 – that is, a total of four free parameters, keeping in mind that the relative inclination between components must be small to allow a single broad shadow to be cast(as opposed to two narrow shadow lanes). The inclination and position angle of zone 4 is set to the values provided in Sect. 3.
The grid was sampled in steps of 1$^{\circ}$ for inclination, and 2$^{\circ}$ for PA, and later refined to 0.5$^{\circ}$ steps for inclination and 0.5$^{\circ}$ for PA after a good initial agreement is found between the average azimuthal profile of the model and the scattered light images. The best fitting model was picked not only based on the location of the shadows cast by the inner components on the outer disk, but also on the shape (slope) of the resulting azimuthal profile of the outer disk. There is also a degeneracy if we consider that the inclinations and PAs of the two inner components (zones 1-2 and zone 3) can be exchanged and produce very similar results. However, doing this would cast a shadow in a different location, and thus produce a different azimuthal profile, for zone 3.

While it is possible to obtain the same degree of misalignment between components by simultaneously changing one of the components' inclination and PA, doing so shifts the location of the shadow. Conversely, it is possible to maintain the location of the shadow by changing the inclination and PA of one of the components in a correlated way.  However, doing so will change the degree of misalignment. Changing the misalignment between components changes the depth of the shadow and the slope of the peak in the azimuthal profile of the outer disk.

For each model, the temperature structure  was obtained with \texttt{MCMax3D}, and subsequently the code raytracing module was used to produce polarized images of each model at both 1.2\,$\mu$m and 1.6\,$\mu$m. In order to compare the model to our combined image, we simulated SPHERE observations through the following procedure: the Stokes Q and U images of the model were convolved with a 2D Gaussian kernel of FWHM=4 pixels ($\sim$0.05”) and 3 pixels ($\sim$0.037”) for the H- and J-band, respectively, to take into account the resolution of the observations (considered at diffraction limit); the coronograph 2D transmission profiles (Wilby et al. in prep) were added at the center of each Q and U image; the H- and J-band model images were then combined in the same way as the data, in a 1:2 ratio to account for the one H-band and two J-band epochs; the convolved Q and U images were combined to obtain Q${\phi}$, U$\phi$ images for each band \citep[see e.g.,][]{deboer2016}. The combined model image was scaled by a constant factor so that the peak flux at the radius of the bright ring at $\sim$16\,au matches that of our combined dataset, and noise was measured from the U${\phi}$ combined scattered light image at every pixel, and then added randomly to the model at each pixel following a normal distribution.

\begin{figure*}[!p]
\center
\includegraphics[width=0.75\textwidth]{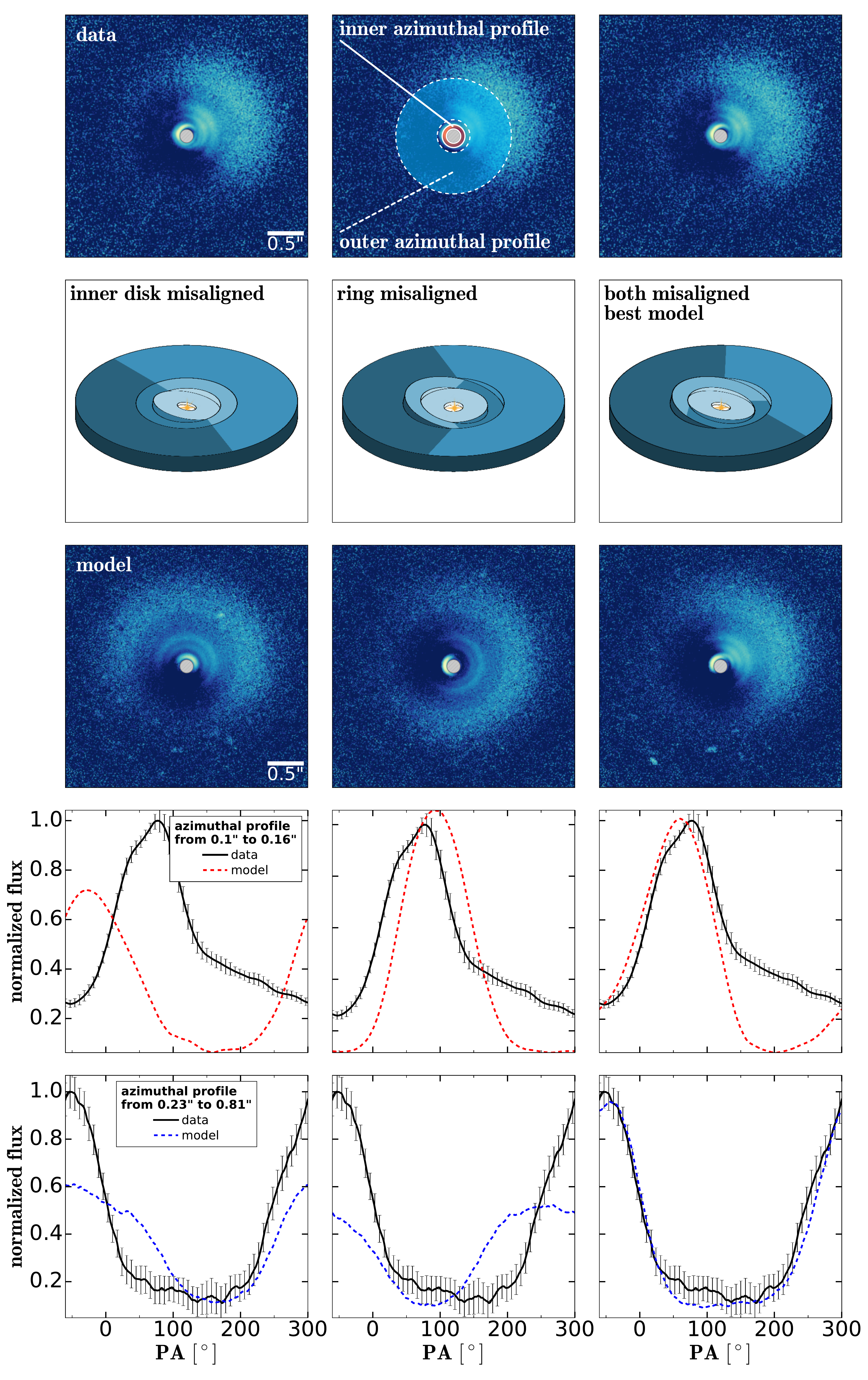}
\caption{
SPHERE combined data set (first row) compared with three radiative transfer models. The second row indicates three configurations of the different disk zones and the corresponding radiative transfer model predictions are shown in the third row. The fourth and fifth rows show the (radially averaged) azimuthal profiles,  from 0.1\arcsec{} to 0.16\arcsec{}, and from 0.23\arcsec{} to 0.81\arcsec{}, respectively. The middle top panel indicates the regions over which the azimuthal profiles were calculated (red + solid white outline, zone 3; blue + dashed white outline, zone 4). 
\textit{Left:} Model including a misalignment of zones 1-2.
\textit{Middle:} Model with a misalignment of zone 3.
\textit{Right:} Our best model with both zones 1-2 and 3 misaligned.}
\label{fig:matrix}
\end{figure*}

\begin{figure}[t]
\center
\includegraphics[width=0.5\textwidth]{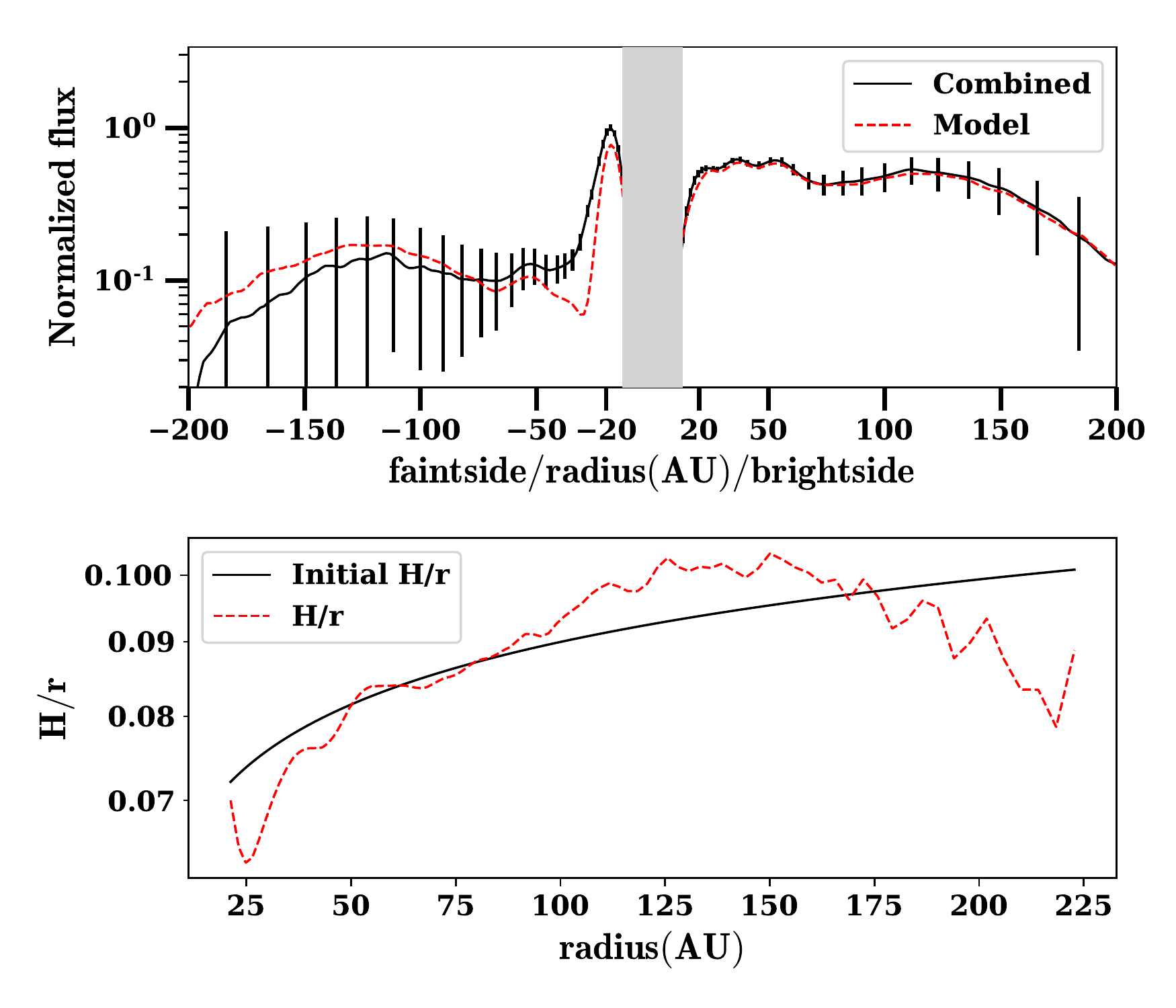} 
\caption{\textit{Top:} average radial profiles for the combined data (black, error bars), and for our model (red, dashed) after the iterative scale height modeling. \textit{Bottom:} aspect ratio of zone 4 before (black, solid) and after (red, dashed) the iterative scale height modeling. } 
\label{fig:scaleheight}
\end{figure}

The azimuthal profiles were obtained for the resulting combined image of each model in our grid, at multiple radii spanning both the bright ring around the coronograph (zone 3) and the asymmetrical outer disk (zone 4). The best of these models was then selected based on the resulting average azimuthal profile of the outer disk (averaged over radii of 30 to 150\,au). Our best model, along with two additional illustrative models, are shown in Fig.\,\ref{fig:matrix}. The first row shows our combined data, the second row presents three sketches representative of different  radiative transfer models shown in the third row. From left to right in the third row, we present the model with only the inner components (zone 1 and 2) misaligned (left column); the model with only the bright ring (zone 3) misaligned (middle column); the model with all 3 zones (zones 1-2 and 3) misaligned with respect to the outer disk (right column; our best model). The fourth and fifth rows show
the average azimuthal profiles of these models computed in the inner disk (from 0.1\arcsec{} to 0.16\arcsec{}) and outer disk (from 0.23\arcsec{} to 0.81\arcsec{}), respectively.  
We clearly see in the fifth row that the misalignment of a single individual component (zones 1-2 or zone 3) is unable of producing a shadow spanning the broad range of position angles as the one observed in our images. 
The superposition of both shadows, however, yields a good fit to both the radially-averaged azimuthal profile shown in the fifth row, as well as to the individual azimuthal profiles measured at different radii in the outer disk (see Fig. \ref{fig:azimuthal_all}), both in terms of the width and overall shape/slopes of the brightness peak. Additionally, the brightness of the shadowed side of the outer disk increases at a radius of $\sim$100\,au in our 
model, just as observed in the SPHERE/IRDIS data, as seen in Fig.\,\ref{fig:scaleheight}, top panel. We confirm that in our model this is due to the outer disk pulling up above the combined shadow of the two inner components at this location as the aspect ratio of the disk increases towards larger radii, as initially predicted.

The fourth row of Fig.\,\ref{fig:matrix} shows the azimuthal profiles of the ring (zone 3) averaged over a radius of 0.1-0.16\arcsec{} ($\sim$14-21\,au). The model in which only the middle component (ring, zone 3) has been misaligned (middle column) shows a relatively good agreement between the azimuthal ring profile of the model with the data, while the model in which only the inner component (zones 1-2) has been misaligned (left column) shows a poor agreement, with the location of the peak brightness of the ring off by over 100 deg. The best agreement, however, is obtained for the model with both components misaligned (right column), with the location, amplitude and shape of the brightness peak of the model in agreement with our data between PAs of ~300 to 120$^{\circ}$. The apparent missing flux in our model between angles of 120 to 300$^{\circ}$ will be discussed in Sect.\,\ref{sec:discussion}. For our best model, the derived inclinations and PAs, as well as the relative misalignment between components, are given in Table\,\ref{tab:misalignments}.

\begin{table}[t!]
 \centering
 \caption{Inclinations and PAs of the different disk components, as well as misalignment $\beta$ with respect to zones 3 and 4.}
  \begin{tabular}{@{}lcccc@{}}
  \hline 
 Zone  		& $i$ ($^{\circ}$)	& PA ($^{\circ}$) & $ \beta_{\rm{zone}-3}$ ($^{\circ}$)& $\beta_{\rm{zone}-4}$ ($^{\circ}$)\\
 \hline 
Zones 1-2  & 20.6 & 272 & 4.8 & 3.3\\ 
Zone 3 & 17.6 & 260.5 &  - & 4.8\\
Zone 4 & 17.6 & 276.5 &  4.8 & -\\
 \hline

\hline\end{tabular}
\label{tab:misalignments}
\end{table}

\begin{figure}[t]
\includegraphics[width=0.48\textwidth]{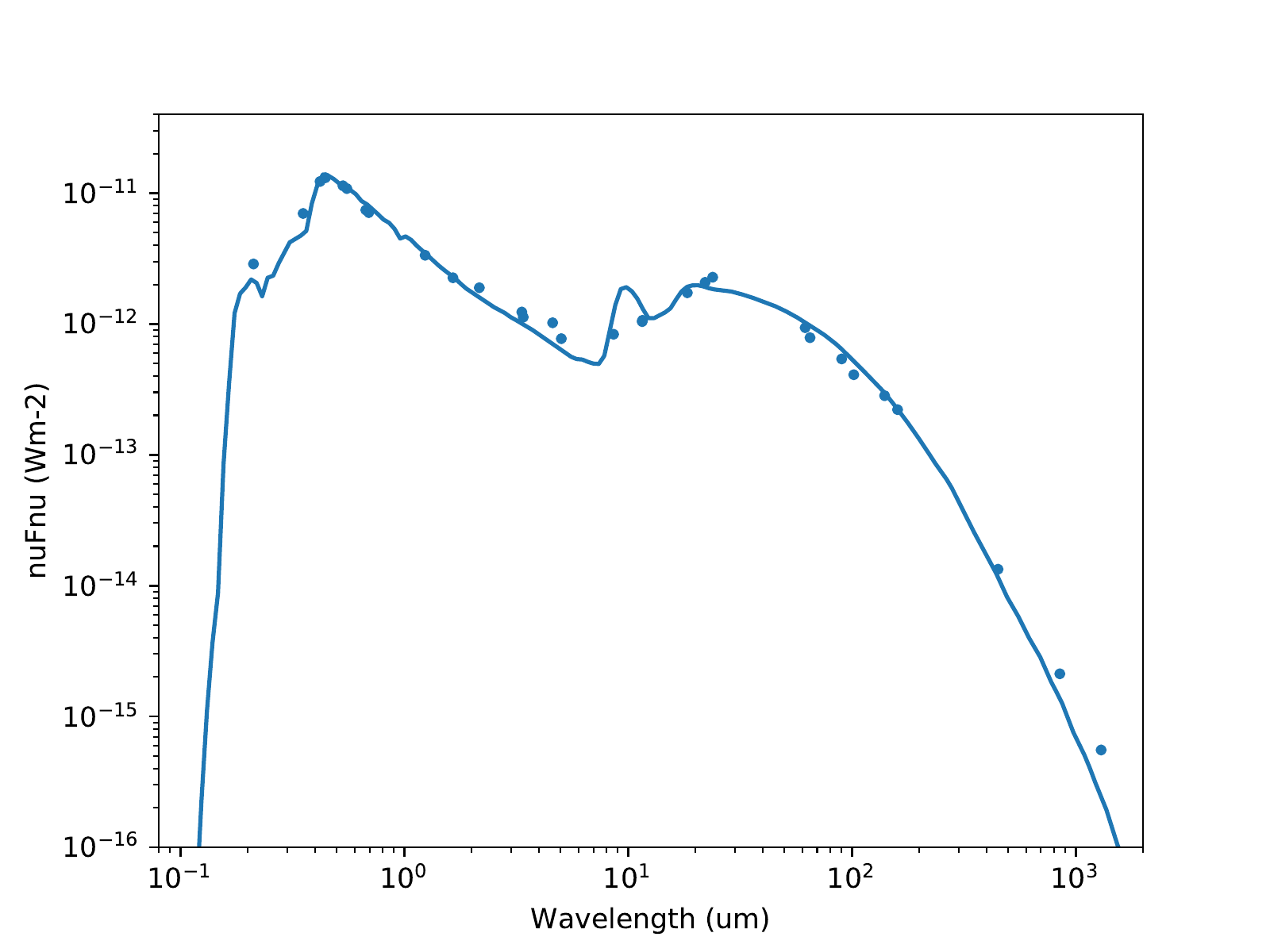} 
\caption{
Spectral energy distribution of HD\,139614 (dots) shown against our radiative transfer model prediction (full line).}
\label{fig:SED}
\end{figure}

In order to simulate Arcs 1 and 2 in the asymmetrical outer disk, we followed an iterative procedure to model the scale height of zone 4 similar to the iterative method used to model the surface density of HD\,163296 in \citet{muro-arena-2018}. The ratio between the radial profiles of model and observations, obtained after azimuthally averaging over position angles of 240 and 360$^{\circ}$, is used to scale the scale height of zone 4 in each iteration until a good agreement between the observed and model radial profiles is obtained. The radial profile was only fit to the bright side of the disk, since Arcs 1 and 2 are not meaningfully visible on the shadowed side of the disk. Figure\,\ref{fig:scaleheight} shows the initial and final scale height profiles for zone 4 (bottom panel), as well as the azimuthally-averaged radial profiles of the combined dataset and of the model after modeling the scale height profile (top panel). This fit was not weighted by the uncertainties in the radial polarized intensity profile, and it was assumed that the radial scale height profile is azimuthally symmetric. Figure\,\ref{fig:SED} shows that our best model provides a qualitative agreement with the SED.

One final feature present in all coronographic epochs and not discussed so far, is the faint Arc 3 feature indicated in the annotated bottom-right panel of Fig. \ref{fig:all_epochs}, along the South-West side of the disk. A similar feature arises naturally in the images of our model, albeit on the North-West side. In our model, this is not a feature in the dust density distribution of the disk, but rather the inner wall of the outer disk (r$\sim$21 au) projected on the plane of view at the outer disk inclination of 17.6 deg. We consider reproducing this feature beyond the scope of this paper, and therefore we do not attempt to fit it. 

\section{Discussion}
\label{sec:discussion}

\subsection{Origin of the misalignments}
A number of observations of protoplanetary disks at different wavelengths support the idea that various disk regions could be misaligned. Scattered light images show dark regions indicative of a large misalignment between inner disk material and outer disk in HD\,142527, HD\,100453, SAO\,206462, and RXJ1604  \citep{marino2015,benisty2017,stolker2016, pinilla2015}, and of a moderate one in HD\,143006 \citep{benisty2018} and DoAr\,44 \citep{casassus2018}. Non-Keplerian gas motion as observed with ALMA are suggestive of warped disk regions, as in AA\,Tau \citep{loomis2017}. Fast changes (timescales $\sim$days) in the properties of the shadows (location, shape) of SAO\,206462 \citep{stolker2017} and RXJ1604 \citep{pinilla2018c} indicate that the innermost regions of these disks are highly dynamic, or that irregular accretion might play a role.

The presence of a massive companion in the disk can induce a misalignment between disk regions \citep[e.g.,][]{facchini2013}. While the presence of a low mass stellar companion can not be excluded in disk regions not accessible by direct imaging, a massive inclined planet can also lead to a significant misalignment of the inner disk, if the planet angular momentum is larger than that of the inner disk \citep[e.g.,][]{matsakos2017}. In that case, the inner disk can break from the outer disk and precess independently from the outer disk \citep[e.g.,][]{bitsch2013,xg2013,nealon2018}. \citet{zhu2019} studied the effect of an inclined planet, held on a fixed orbit, on the depth and width of the gap and show that the breaking happens under the condition that the gap carved by the planet is deep enough or that the disk viscosity is very low. Alternatively, secular precession resonances can lead to large misalignments between the inner and outer disks if the companion is quite massive and no viscosity is considered \citep[0.1-0.01\,M$_\odot$,][]{owen2017}. In the case of the transition disk HD\,142527, that shows two shadows in scattered light images, a stellar companion was detected \citep[0.2-0.3\,M$_\odot$,][]{biller2012}, likely on an inclined and eccentric orbit \citep{lacour2016,claudi2019}, which would explain the misalignment of the inner disk \citep{price2018}. However, apart from HD\,142527, none of the disks with scattered light shadows are known to host a massive inclined companion. 

In the case of HD\,139614, our modeling implies that the relative misalignments between the inner components (zone 1-2) and the bright ring (zone 3), and the outer disk (zone 4), are quite small (<10 deg), which might also be consistent with the presence of low mass planets. While the disk breaking does not occur with such a low mass planet and its inclination is damped rapidly \citep{bitsch2011}, it can still warp the disk when its inclination is $\sim$2-3 times the disk aspect ratio at the planet location when held on a fixed orbit \citep{Arzamasskiy2018}. \citet{juhasz2017} provided observational diagnostics of such small warps and show that even in these cases, the self-shadowing of the disk results in surface brightness asymmetries that are clearly detectable in scattered light images, as well as kinematical asymmetries in gas lines. Similar findings are obtained by \citet{nealon2019} who find that even a tiny misalignment (less than 1$^\circ$) between the inner and outer disk, due to a few Jupiter mass planet, leads to shadowing of the outer disk. 

While our scattered light images are very sensitive to the illumination of the outer disk, they are not suitable to find planets or low mass stellar companions, since they were carried out in polarimetric mode. Constraints on the mass of a companion are therefore not possible based on these observations alone. The previously mentioned works however suggest that such a companion would be located inside of the misaligned region, which we cannot trace with direct imaging. We can only speculate that this companion could be located inside of $\sim$6\,au, which is a region that must be dust-depleted in order to reproduce the lack of IR excess in the SED at the corresponding wavelengths.

The possibility of multiple companions being responsible for the disk misalignment or warp, as well as for the arcs in the outer disk, cannot be discarded. Since rings in scattered light images likely trace small variations of the slope of the scattering surface, rather than density variations, observations of the disk midplane, for example with ALMA, would be needed to determine if these features have mm counterparts. Until then, claims on the number of companions in this disk are speculative. HD\,139614 is therefore a prime target for a high resolution study with ALMA in the continuum, but also to look at kinematical evidence for a warped region.

Interestingly, \citet{garufi2018} note that disks with narrow shadows (hence, large misalignments) also have a very high near IR excess ($\sim$25\%) likely indicating a large vertical extent for the inner disk. In contrast, HD\,139614 has a low near IR excess ($\sim$8\%) and our modeling work indicates a depleted inner disk and  small misalignments between disk regions. It is therefore possible that the amount of near IR excess directly relates to the amount of vertical stirring by an inclined companion, responsible for the misalignment. 

Finally, we note that while a misaligned magnetic field can warp the inner disk \citep{bouvier2007}, it is unlikely to be the cause of the large scale shadow in HD\,139614, as Herbig stars do no possess strong magnetic fields \citep{alecian2013}.

\subsection{Number of misaligned components} 
All epochs show an outer disk with asymmetrical brightness  ($\sim$30-220\,au) 
that, when attributed to shadowing, cannot be explained by a single misaligned component due to the broad angular extent of the shadow. This is seen clearly in Fig. \ref{fig:matrix}, where the effects of inclining two different single components in our model can be seen in the left and middle columns; in both cases the shadow cast by each individual component appears too narrow, and increasing the inclination of the misaligned component only serves to produce two narrow shadowed lanes on the outer disk rather than making this single shadow broader. Similarly, the asymmetry observed in the ring at $\sim$16 au requires an inner component to be misaligned. The fact that the peaks of the azimuthal profiles of the ring and outer (r>30\,au) disk are offset by $\sim$120 deg rather than 180 deg supports this hypothesis.

The inclination and PA of these components can be well-constrained not just by the location of the shadows they cast, but also by the slope/shape of the azimuthal profile peak of both the outer disk and the ring. While it is possible to produce a very similar azimuthal profile of the outer disk by interchanging the inclinations and PAs of both inner components, only one of these two configurations locates the peak of the azimuthal profile of the ring at the correct location. 
Furthermore, while there is some inclination-PA degeneracy for each of the inner components when estimating them using the method described in Sect.\,\ref{sec:modeling}, we stress that larger misalignments will easily produce dual shadowed lanes as opposed to single broad shadows on the outer components, so we can safely conclude that the misalignment between components should be small and of the order of <10 deg.

\begin{figure}[t]
\center
\includegraphics[width=0.5\textwidth]{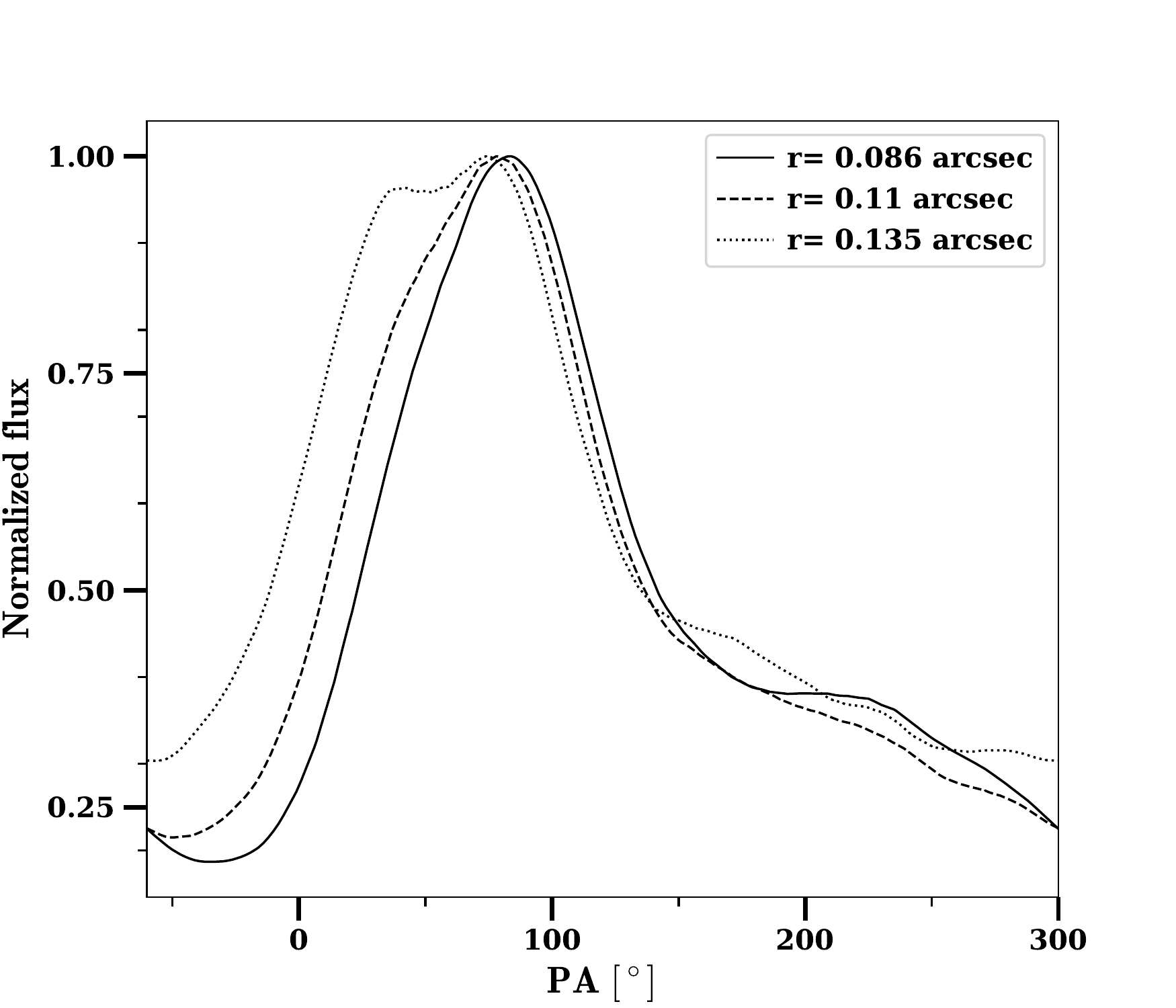} 
\caption{Azimuthal profiles of the data at three different radii across the location of the ring feature: the peak of the brightness profile/location of the shadow can be seen shifting towards smaller position angles with increasing radius.} 
\label{fig:warp}
\end{figure}

Whether the ring and the inner/outer disks are joined by a continuously-warped region is difficult to determine from our polarized scattered light images. The apparent gap observed in the H-band non-coronagraphic epoch at r$\sim$11 au could be the product of shadowing rather than a real gap in the disk dust density distribution. Between the ring and the outer disk, on the other hand, there is no gap apparent in the data at the instrument resolution of $\sim$50\,mas (H-band, $\sim$7 au) and ~40 mas (J-band, $\sim$5.5 au). If this region is warped, we expect this to cause an azimuthal shift of the shadow of the innermost component with increasing radius. This is consistent with what we see in our data if we analyze in detail how the azimuthal profile of the ring changes with radius. Fig.\,\ref{fig:warp} shows the radial variation of the azimuthal profile of the disk between radii of $\sim$0.086 and 0.135\arcsec{}, with the peak flux of the ring shifting towards smaller PAs (between $\sim$88 and 75$^{\circ}$) over this small radial extent. So while we modeled this region as two separate components (a ring between 14 and 21 au and an outer disk between 21 and 220\,au), it is quite possible that these are joined by a continuous warped region. This might help explain the missing flux we observe in the azimuthal profiles of the ring between angles of $\sim$120 and 300 deg in PA in the fourth row of Fig. \ref{fig:matrix}. 

Hydrodynamical simulations of warps in disks show a continuous warping of the disk exterior to the companion. \citet{nealon2019} find that the combined effect of shadowing from the disk interior to the companion, plus the shadowing caused by the warp in the outer disk, leads to azimuthal variations that decrease in amplitude with increasing radius. This is consistent with the radial behavior of the azimuthal variations seen in the outer disk of HD\,139614. \citet{juhasz2017} show that the warp created by an inclined equal mass binary companion can lead to a broad (>180$^{\circ}$) shadow on the outer disk (see their Fig.\,5), although not as broad as our observations require.  However, models with such a massive companion might not be appropriate for HD\,139614, and the shape of the shadow is strongly dependent on the exact morphology of the warp, in particular on its radial extent. This extent is also time dependent, as the companion tries to align the disk with its orbital plane. Therefore, although we acknowledge that instead of two misaligned rings, the system might host a single continuous warp, determining its  exact shape requires extensive hydrodynamical modeling, which is beyond the scope of this paper.

\subsection{Similarity to HD\,100546}

The mid-IR SED and 10$\mu$m silicate emission spectrum of HD\,139614 shows a strong similarity to those of transitional disk HD\,100546. HD\,100546 shows a pronounced crystalline silicate emission feature first observed by \citet{bouwman2003}, who propose that this feature is connected to the presence of a possible giant planet carving a gap in the disk. These crystalline silicates could be formed locally in the outer disk in planet-induced shocks or violent collisions between planetesimals dynamically excited by the planet. The location of these silicates (forsterite) was later shown to coincide with the inner wall of the outer disk at $\sim$15 au from the star by \citet{mulders2011}. The 10$\mu m$ spectrum of HD\,139614 obtained with TIMMI2 at the ESO 3.6m telescope \citep{vanboekel2005} and with Spitzer \citep{juhasz2010} suggests that most of the silicate emission from this disk arises from a region corresponding to somewhere between the outer edges of zone 2 and the inner edge of zone 4 in our model. If this emission indeed originates from planet-induced shocks or planetesimal collisions, as theorized by \citet{bouwman2003} and supported by the location of the gap as determined by \citet{mulders2011}, it would be consistent with the location of a hypothetical planet causing the warp or misalignment between components observed in the disk.

\section{Conclusions}
Shadowing by different disk components in scattered light imaging is a powerful tool to infer the inner disk morphology. Several transitional disks have been shown to contain undetected or unresolved inner disks with high inclination, causing a pair of shadows on the inner rim of the outer disk and actually the entire outer disk \citep{avenhaus2014,benisty2017,Min2017,marino2015}.  Such shadows can be strongly time-variable \citep{stolker2017,pinilla2018c}.  In this paper we discussed a different case in which the relative inclination between different disk parts is only small, leading to very broad, one-sided shadow features.  We used the observations of HD139614 with VLT/SPHERE along with radiative transfer modeling and arrived at the following conclusions:

\begin{enumerate}
    \item The image of HD\,139614 shows a very broad shadow region in the outer disk (30-200 au), between approximate position angles 0$^{\circ}$ and 240$^{\circ}$.
    \item Such a shadow cannot be produced by a single inclined ring component, because the shadow of  a component like that would be at most $\sim$180 degrees wide. A continuously warped region could also lead to such a wide shadow.
    \item The ring around 0.12\arcsec{} shows a nearly almost opposite behavior compared to the outer disk, being bright between position angles $\sim$0 and 130$^{\circ}$ and dark between position angles $\sim$130 and 360$^{\circ}$.
    \item The observed asymmetries can be reproduced with a model that contains two ring-shaped disk parts that are inclined with respect to each other by $\sim$4$^{\circ}$ and also with respect to the outer disk. The two components might also be connected by a warp. This geometry gives a good fit to the azimuthal profiles of much of the disk. The overall SED of HD\,139614 is also well-reproduced by our model.
    \item A warped disk and discrete misaligned regions can be caused by either a single or multiple companions in misaligned orbits. Since we can not distinguish between these scenarios at this point we cannot constrain the number of possible companions in the disk. If the small misalignments are tracing weakly-inclined planets perturbing the disk, this system suggests that planets could be mutually misaligned already during the formation phase. 
    \item The outer disk contains additional ring structures that can be well described by a modulation of the disk scale height.
\end{enumerate}


\section*{Acknowledgements}
We thank the anonymous referee who helped improved our manuscript with constructive comments. M.B. acknowledges A.~Juh\'asz for insightful discussions. SPHERE is an instrument designed and built by a consortium
consisting of IPAG (Grenoble, France), MPIA (Heidelberg, Germany),
LAM (Marseille, France), LESIA (Paris, France), Laboratoire Lagrange
(Nice, France), INAF - Osservatorio di Padova (Italy), Observatoire de
Genève (Switzerland), ETH Zurich (Switzerland), NOVA (Netherlands), ONERA
(France), and ASTRON (The Netherlands) in collaboration with ESO.
SPHERE was funded by ESO, with additional contributions from CNRS
(France), MPIA (Germany), INAF (Italy), FINES (Switzerland), and NOVA
(The Netherlands). SPHERE also received funding from the European Commission
Sixth and Seventh Framework Programmes as part of the Optical Infrared
Coordination Network for Astronomy (OPTICON) under grant number RII3-Ct2004-001566
for FP6 (2004-2008), grant number 226604 for FP7 (2009-2012),
and grant number 312430 for FP7 (2013-2016). G.M-A. acknowledges funding from the Netherlands Organisation for Scientific Research (NWO) TOP-1 grant as part
of the research program “Herbig Ae/Be stars, Rosetta stones for understanding
the formation of planetary systems”, project number 614.001.552. M.B.  acknowledges funding from ANR of France under contract number ANR-16-CE31-0013 (Planet Forming disks). A.Z. acknowledges support from the CONICYT + PAI/ Convocatoria nacional subvenci\'on a la instalaci\'on en la academia, convocatoria 2017 + Folio PAI77170087. 
The  figures were generated with the \textsc{python}-based package \textsc{matplotlib} \citep{matplotlib}.

\bibliographystyle{aa}
\bibliography{hd139_v0} 

\begin{appendix}
\section{Polar projection}
\label{app:CO}

\begin{figure}[th]
\includegraphics[width=0.47\textwidth]{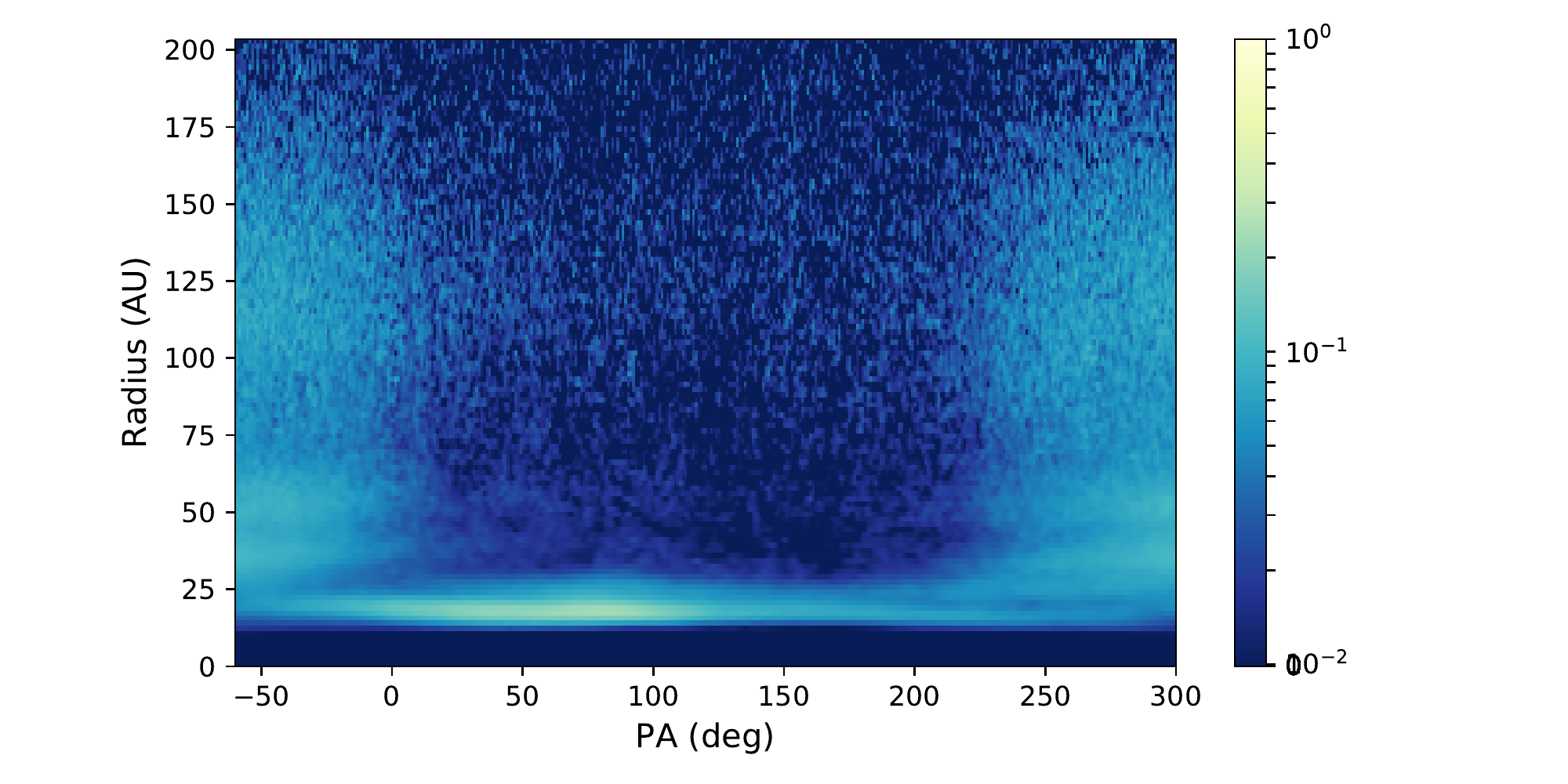} 
\caption{Polar projection of the combined dataset: the shadowed region is visible in the center of the figure for radii larger than 30 au, constrained to an approximately constant azimuthal range. The bright ring can be seen at r$\sim$15 au, with its brightest region partially overlapping with the darker region of the outer disk.
} 
\label{fig:polar_projection}
\end{figure}

Figure \ref{fig:polar_projection} shows the full polar projection of the disk observations, which shows dark and bright regions in both the inner and outer disks and illustrates the opposite behavior as a function of position angle between these regions.  

\section{Azimuthal profiles}
\begin{figure}[th]
\includegraphics[width=0.5\textwidth]{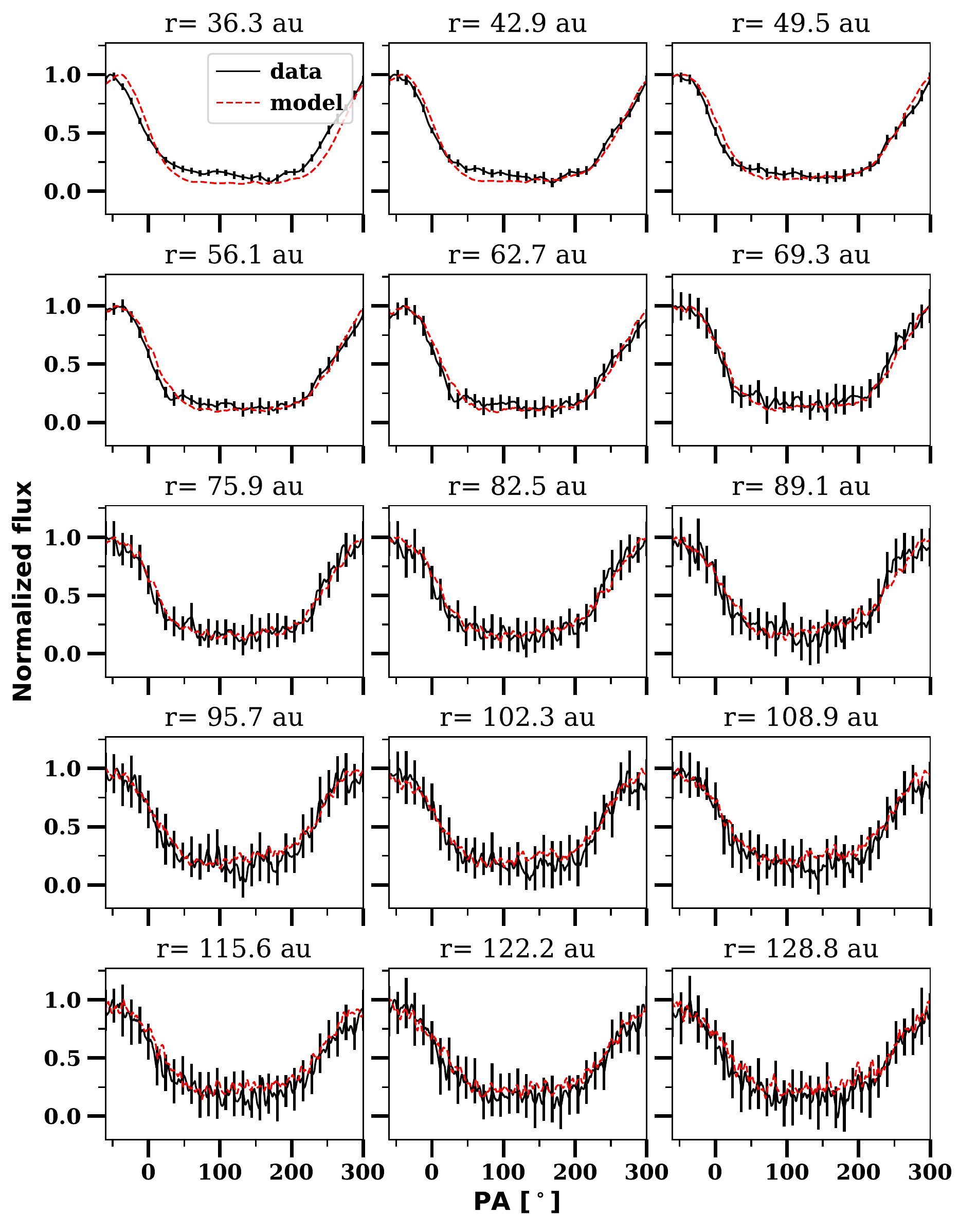} 
\caption{Azimuthal profiles of the outer disk for both the combined dataset (black) and our model (red) at different radii. Each profile was measured in an annulus of width 4 pixels ($\sim$6.6 au) centered on the radius indicated on each subfigure.} 
\label{fig:azimuthal_all}
\end{figure}

Figure \ref{fig:azimuthal_all} shows the excellent fit of the brightness profiles between model and observations at all radii.

\end{appendix}

\label{lastpage}
\end{document}